\documentstyle[aps,prl,epsf,twocolumn,floats]{revtex}

     \def\d{\delta}
     \def\e{\varepsilon}

     \def\@cite#1{{\footnotesize $^{#1}$}}
\begin{document}
\draft
\twocolumn[\hsize\textwidth\columnwidth\hsize\csname @twocolumnfalse\endcsname
\title{Localization-induced Griffiths phase of disordered Anderson
lattices}
\author{E. Miranda$^{\, (1)}$ and V. Dobrosavljevi\'c$^{\, (2)}$}
\address{$^{(1)}$Instituto de F\'{\i}sica ``Gleb Wataghin'', Unicamp,
C.P. 6165, Campinas, SP, CEP 13083-970, Brazil.\\
$^{(2)}$Department of Physics and 
National High Magnetic Field Laboratory,
Florida State University, 
Tallahassee, FL 32306.}
\maketitle

\begin{abstract}
We demonstrate that local density of states fluctuations in disordered
Anderson lattice models universally lead to the emergence of non-Fermi
liquid (NFL) behavior.  The NFL regime appears at moderate disorder
($W = W_c$) and is characterized by {\em power-law} anomalies, e. g.
$C/T \sim 1/T^{(1-\alpha )}$, where the exponent $\alpha$ varies
continuously with disorder, as in other Griffiths phases.  This
Griffiths phase is {\em not} associated with the proximity to any
magnetic ordering, but reflects the approach to a disorder-driven
metal-insulator transition (MIT).  Remarkably, the MIT takes place
only at {\em much} larger disorder $W_{MIT} \approx 12 W_c$, resulting
in an extraordinarily robust NFL metallic phase.
\end{abstract}

]
\narrowtext

The nature of the non-Fermi liquid (NFL) behavior observed in several
heavy fermion compounds remains largely unresolved\cite{general}. In
the cleaner systems, the proximity to a quantum critical point seems
to be at the origin of many of the observed properties
\cite{qcp-ex,qcp-th}. Exotic impurity models cannot be
discarded\cite{dan}, though their behavior in concentrated systems
remains ill understood\cite{jarrell}.

In other compounds, non-stoichiometry has prompted the investigation
of disorder-based mechanisms.  A {\em phenomenological} ``Kondo
disorder'' model (KDM), describing a broad distribution of Kondo
temperatures $T_K$, has been successfully applied to several of these
systems\cite{ourprev,kondodis}.  Alternatively, the formation of large
clusters of magnetically ordered material within the disordered phase
has also been proposed\cite{antonio}. Both scenarios lead to a wide
distribution of energy scales, giving rise to similar thermodynamic
anomalies and NMR response\cite{kondodis}.  In addition, the
predictions of the KDM prove to be consistent with a number of other
experiments, including optical conductivity\cite{optics},
magnetoresistance\cite{magnet} and dynamic neutron scattering
\cite{ourprev,neutron} measurements.

Despite these successes, a number of basic questions remain
unresolved, including: (1) What is the {\em microscopic} origin of the
ubiquitous power law (or logarithmic) anomalies?  (2) Can a model
calculation be done, which can produce these power laws in a universal
fashion? (3) Are these properties tied to the proximity to a quantum
phase transition, and if so, which one?  (4) How robust is the
anomalous behavior with respect to the variation of materials
parameters?

Within our model, all these questions find clear-cut and physically
transparent answers: (i) The anomalies can be ascribed to a {\em power
law} distribution of $T_K$'s, whose exponent varies continuously with
disorder strength. The resulting NFL behavior, e. g. $\gamma =C/T
\sim 1/T^{(1-\alpha )}$, $\alpha < 1$ sets in for relatively weak
randomness, {\em irrespective} of the detailed model for disorder.
This should be contrasted with the KDM \cite{ourprev,kondodis}, where
the occurence of NFL behavior requires fine-tuning.  (ii) We find
universal behavior reflecting the {\em nonlocal}, many-body processes
associated with Anderson localization effects in the presence of
strong electron correlations.  (iii) for stronger disorder, the NFL
metallic behavior persists over a surprisingly large interval before a
disorder-driven MIT is reached.  This novel Griffiths phase is a
manifestation of quantum critical behavior associated with the
approach to a disorder-driven metal-insulator transition and does not
require the proximity of any magnetically ordered phase.

We consider a disordered infinite-U Anderson lattice Hamiltonian
\begin{eqnarray}
H &=& \sum\limits_{ij\sigma} \left(-t_{ij} + \e_i\d_{ij}\right)
c^{\dagger}_{i\sigma} c^{\phantom{{\dagger}}}_{j\sigma}
+ \sum\limits_{j\sigma} E_{fj} f^{\dagger}_{j\sigma}
f^{\phantom{{\dagger}}}_{j\sigma} \nonumber \\
&+& \sum\limits_{j\sigma} V_j (c^{\dagger}_{j\sigma}
f^{\phantom{{\dagger}}}_{j\sigma}  + {\rm H. c.} ),
\label{hammy}
\end{eqnarray}
in usual notation.
The infinite-U constraint at each f-orbital is assumed ($n^f_j
=\sum_{\sigma}f^{\dagger}_{j\sigma} f^{\phantom{{\dagger}}}_{j\sigma}
\le 1$). We have studied different types of disorder, including
randomness in the conduction electron site energies $\e_i$, the
f-electron energies $E_{fj}$, as well as the hybridization
$V_j$. Within our approach, we find that most of our conclusions
remain valid for {\em any} specific form of disorder, indicating
robust and universal behavior.

We treat the above Hamiltonian within the recently proposed
statistical dynamical mean field theory\cite{sdmft}.  This approach
reduces to the usual dynamical mean field theory in the limit
$z\rightarrow\infty$ (with $t_{jk}\sim
t/\sqrt{z}$)\cite{ourprev,dinf,dinfdis}, but unlike the latter, it
incorporates Anderson localization effects.  As a result, the spectral
function of the local bath ``seen'' by each impurity has strong
spatial fluctuations and contains information coming from sites which
are many lattice parameters away.  Physically, the fluctuations of the
conduction electron wave-functions lead to the distribution of Kondo
temperatures, which in turn creates a {\em renormalized} effective
disorder seen by the conduction electrons. This nonlocal feedback
mechanism results in the universal form of all the relevant
distribution functions that we find.

The simplest model for incorporating localization effects is obtained
by focusing on a Bethe lattice of coordination $z$ (with nearest
neighbor hopping $t$, used as unit of energy). The resulting set of
self-consistent stochastic equations is governed by the local
actions\cite{sdmft,sces}
\begin{eqnarray}
S_{{\rm eff}}^{(j)} &=&\int_{0}^{\beta }d\tau 
\sum_{\sigma }f_{j,\sigma }^{\dagger }(\tau )\left( \partial _{\tau
}+E_{fj}\right)f_{j,\sigma
}(\tau ) +\nonumber \\
&&\int_{0}^{\beta }d\tau \int_0^{\beta}d\tau ^{\prime }
\sum_{\sigma }f_{j,\sigma }^{\dagger }(\tau )
\Delta _{j}(\tau -\tau ^{\prime })f_{j,\sigma
}(\tau ^{\prime });
\label{seff} \\
&&\Delta _{j}(\omega ) = \frac{V_{j}^{2}}{\omega -\e
_{j}-\sum_{k=1}^{z-1}t_{jk}^{2}G_{ck}^{(j)}(\omega )}.  \label{hybrid}
\end{eqnarray}
Here, $G_{ck}^{(j)}(\omega )$ is the local c-electron Green's function
on site $k$ with the nearest neighbor site $j$ removed. It is
determined recursively from\cite{sdmft,sces}
\begin{eqnarray}
G_{cj}^{(i)(-1)}(\omega ) &=&\omega -\e
_{j}-\sum_{k=1}^{z-1}t_{jk}^{2}G_{ck}^{(j)}(\omega ) 
- \Phi_j (\omega ),\\
\Phi_j (\omega ) 
&=&\frac{V_{j}^{2}}{\omega -E_{fj}-\Sigma _{fj}(\omega )}. \label{recur}
\end{eqnarray}
The local self-energy $\Sigma _{fj}{(\omega )} $ is obtained from the
solution of the effective action (\ref {seff})\cite{sdmft,sces,tk}.
In order to solve the impurity problems, we have used the large-N
mean-field theory at $T=0$\cite{largeN}. We have solved
Eqs.~({\ref{seff}-\ref{recur}) numerically by sampling.  In
implementing this procedure, we have carried out large-scale
simulations for $z=3$, with ensembles containing up to $N_s =200$
sites, and frequency meshes containing up to $N_{omega} = 8,000$
frequencies. The numerical integrations needed to solve the impurity
problems have been done by a combination of spline interpolations and
adaptive quadrature routines. These careful numerics have made it
possible to obtain Kondo temperatures spanning {\em fifteen} orders of
magnitude, which was crucial in order to examine the long tails of the
relevant distribution functions.

One of the greatest advantages of our approach is its ability to focus
on full distribution functions, which is essential for characterizing
any Griffiths phase. Some typical results are presented in
Fig.~\ref{fig1}, where we show the evolution of the distribution of
local Kondo temperatures as a function of disorder, from which one
computes the overall response of the lattice system (See the
discussion in the first ref. of \cite{ourprev}). We find that
(Fig.~\ref{fig1}(a)) the distribution has a universal log-normal form
for weak disorder. We have verified that such a log-normal behavior is
obtained {\em irrespective} of the type and shape of the bare disorder
distribution, as long as it is not too large.

\begin{figure}[htb]
\epsfxsize=3.2in \epsfbox{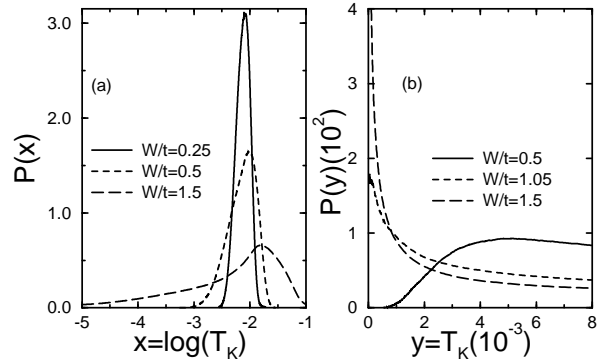}
\caption{(a) Distribution of $\log(T_K)$ for various values of disorder
strength $W/t$. 
 (b) Distribution of $T_K$ showing the emergence of NFL
behavior.  Here, $\e_i$'s are distributed uniformly with width $W$ and
we have used $z=3$, $E_f=-1$, $V=0.5$ and $\mu=-0.5$, in units of $t$.
\label{fig1}}
\end{figure}

As the disorder is increased, the distribution $P(T_K)$ no longer
retains its log-normal form. Instead, a long tail emerges on the
low-$T_K$ side, with a power law asymptotic form (Fig.~\ref{fig1}(b))
\begin{equation}
P(T_K)\sim T_K^{(\alpha-1)}.
\label{power-law}
\end{equation}
The exponent $\alpha$ varies continuously with disorder,
as seen on a plot of $\log(P(\log(T_K)))$ in Fig.~\ref{fig2}.

\begin{figure}[htb]
\epsfxsize=3.2in \epsfbox{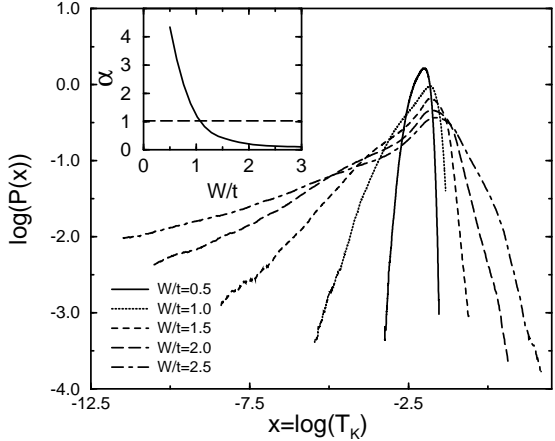}
\caption{Power law asymptotics of $P(\log(T_K))$ as the disorder increases. The
linear behavior for small $\log(T_K)$ implies a power law dependence of
$P(T_K)$. Inset: the exponent $\alpha$ of Eq.~(\ref{power-law}). Same
parameters as in Fig.~\ref{fig1}.\label{fig2}}
\end{figure}

Note that the value $\alpha = 1$, (Figs.~\ref{fig1}(b) and
\ref{fig2}) with $P(T_K)\sim {\rm const.}$, corresponds to the
condition for {\em Marginal} Fermi Liquid behavior observed in some
Kondo alloys\cite{ourprev,kondodis}, with logarithmically divergent
magnetic susceptibility $\chi(T)$ and specific heat coefficient
$\gamma$. This divergent behavior becomes more singular as the
disorder is increased past this marginal case. For example, if we use
the simple Wilson interpolation formula for $\chi(T)$\cite{wilson}
\begin{equation}
\chi(T) \sim \int_0^{\Lambda} \frac{T_K^{(\alpha-1)}}{T+a T_K} dT_K
\sim \frac{1}{T^{(1-\alpha)}}.
\end{equation}

We thus have power law divergences with exponents which vary
continuously with the disorder strength.  If we take $t\sim 10^4 K$,
this should be observed below a few tens of Kelvin. Such generic
behavior has been fitted to some NFL compounds\cite{marcio}. Besides,
$\chi(0)$ will diverge at a critical disorder strength ($W_c \approx
1$ in Fig.~\ref{fig2})
\begin{eqnarray}
\chi(0) &\sim& \int \frac{P(T_K)}{T_K} dT_K \sim \int_0^{\Lambda}
T_K^{(\alpha-2)} dT_K \nonumber \\
&\sim& \frac{1}{\alpha-1} \sim \frac{1}{W_c-W},
\end{eqnarray}
with a similar result for $\gamma$.  Note, however, that other higher
order correlation functions, such as the non-linear susceptibility
$\chi_3(0)$, which probes higher negative moments of the distribution
($\chi_3(0) \sim 1/T_K^3$), will begin to diverge at different
critical values of disorder,
\begin{equation}
\chi_3(0) \sim \frac{1}{\alpha-3} \sim \frac{1}{W_{c3}-W},
\end{equation}
where $W_{c3}\approx 0.66$ for the parameters of Fig.~\ref{fig2}.

This general behavior is characteristic of Griffiths
phases\cite{griffiths} and should not be confused with a true phase
transition.  The system should be viewed as {\em a disordered metal
with embedded clusters of Anderson insulators}. It is precisely these
poor conducting regions, with depleted densities of states, which give
rise to imperfectly quenched spins and the corresponding singular
thermodynamic properties.

We should also stress that the main mechanism that dominates the
Griffiths phase is qualitatively different from the one in the
KDM. There, $T_K$ fluctuations were simply caused by the distribution
of {\em local} parameters ($V_j$, $E_{fj}$) and the conduction
electron DOS does not fluctuate. By contrast, in the present
treatment, fluctuations in the latter are dominant.  To illustrate
this, all the results we present are obtained for a model with
conduction band disorder {\em only}, although similar results follow
for any form of disorder. We stress that, in a KDM treatment of this
case, $T_K$ fluctuations are severely limited.  Here, however, $T_K$
fluctuations are enhanced by the fluctuations in the local conduction
DOS, reflecting the localization effects and the approach to a
disorder-driven MIT.

To confirm this picture, we examine the localization properties of the
conduction electrons.  We focus on the {\em typical} DOS $\rho_{typ} =
\exp\{ < \ln \rho_j >\}$; $\rho_j = (1/\pi )Im G_{cj} (\omega =0)$, as
shown in Fig.~\ref{fig3}. This quantity vanishes at any
disorder-driven MIT\cite{sdmft}, and thus can serve as an order
parameter for localization. Remarkably, we find a strong decrease of
this quantity upon entering the Griffiths phase ($W/t \approx 1$),
reflecting the strongly enhanced conduction electron scattering due to
Kondo disorder. Yet, the actual localization transition, where the
typical DOS vanishes, occurs only at much larger disorder ($W/t
\approx 12$). This results in a very extended NFL metallic region,
where the thermodynamics is singular, and the conduction electrons are
almost, but not completely localized.

\begin{figure}
\epsfxsize=3.2in \epsfbox{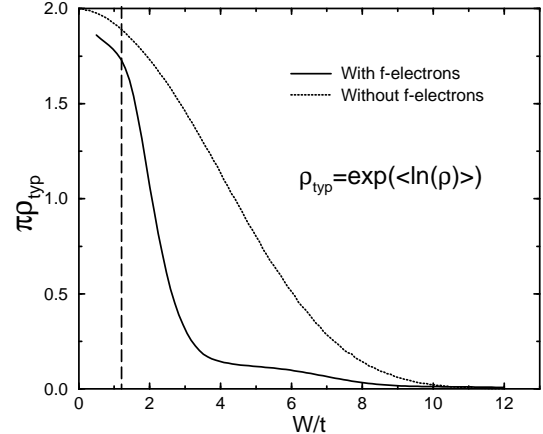}
\caption{Localization properties of the 
conduction electrons as monitored by the typical DOS as a function of
disorder. The same quantity in the absence of Kondo spins is shown for
comparison. The vertical dashed line indicates the boundary of the NFL
phase. Same parameters as in Fig.~\ref{fig1}.
\label{fig3}}
\end{figure}

This dramatic effect has a simple physical origin. Consider the
distribution of the effective scattering potentials of the conduction
electrons $\Phi_j (\omega =0)$ (see Eq. (5)) introduced by the
f-sites. Note that \cite{ourprev} $\Phi_j (\omega =0)= - Z_j
V^2/\tilde{\varepsilon}_{fj}$, where $Z_j$ is the quasiparticle weight
and $\tilde{\varepsilon}_{fj}$ the (renormalized) energy of the Kondo
resonance at site $j$.  For sufficient disorder, the Kondo resonances
are randomly shifted up or down in energy, giving rise to $\Phi_j$'s
that can be random in magnitude but also in {\em sign}.  The resulting
distribution for the inverse quantity $\Phi_j^{-1}$ is shown in
Fig.~\ref{fig4} and is found to broaden with disorder.  For $W/t
\approx 1.5$, a {\em finite density} of $\Phi_j^{-1} =0+$
(i. e. $\Phi_j =+\infty $) sites emerges. This is crucial, since the
corresponding f-sites act as unitary scatterers (US's), characterized
by a maximally allowed scattering phase shift ($\delta = \pi /2$) for
the conduction electrons. If all the f-sites were US's, the system
would be a Kondo insulator. The presence of a finite fraction of US's
should be viewed as the emergence of {\em droplets} of a Kondo
insulator within the heavy metal. Interestingly, at stronger disorder
($W/t > 4$) the distribution of $\Phi_j^{-1}(0)$ continues to broaden,
leading to a {\em decrease} in the number of US's. This is illustrated
by plotting $P(\Phi_j^{-1}=0 )$ in the inset of Fig~\ref{fig4}. In
this regime, while the bare disorder increases, the {\em effective
disorder} produced by the f-sites is reduced, stabilizing the almost
localized metallic phase.  This mechanism may be at the origin of the
puzzling behavior of materials such as $\rm Sm B_6$\cite{aronson},
where the low temperature resistivity remains anomalously large yet
finite over a broad range of parameters.

\begin{figure}
\epsfxsize=3.2in \epsfbox{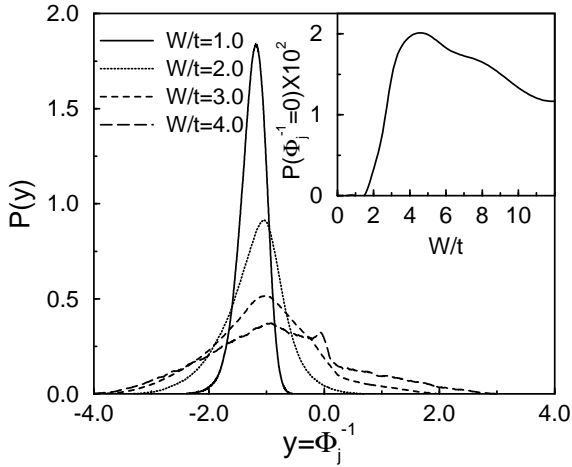}
\caption{Distribution function for the 
quantity $\Phi_j^{-1}(0)$ as a function of disorder.  The inset shows
the behavior of the concentration of the strong scattering Kondo
centers ($\Phi_j^{-1}(0)=0$), which reaches a maximum just after the
NFL phase is entered. Same parameters as in Fig.~\ref{fig1}.
\label{fig4}}
\end{figure}

Finally, we note that a similar NFL phase was identified in a study of
the Mott-Anderson transition in the disordered Hubbard
model\cite{sdmft}.  We have re-examined this system, and concluded
that the NFL behavior should be attributed to a related Griffiths
phase rather than a separate thermodynamic phase of the
system. Despite the similarities, several features prove dramatically
different. For Hubbard models, the emergence of NFL behavior does not
have a dramatic effect on the conduction electrons and no US's emerge.
This observation may explain the strong correlation between
thermodynamic and transport anomalies in Kondo alloys, but not in
doped semiconductors.  In the latter materials, the thermodynamics is
still singular close to the MIT, while transport remains more
conventional\cite{paalanen}.

It would be of particular interest if it could be tested
experimentally whether these localization effects are responsible for
the observed NFL behavior of disordered heavy fermion systems. A
scanning tunneling microscopy study might be able to detect the
predicted insulator droplets. In order to distinguish this from the
magnetic Griffiths phase scenario\cite{antonio}, a systematic study of
systems with comparable amounts of disorder but different magnetic
character would be useful. Besides, since the present theory relies
very little on intersite magnetic correlations, a determination of the
typical size of the relevant magnetic clusters could also serve as a
test.

In summary, we have investigated and solved a microscopic model for
disordered Anderson lattices that displays an unprecedented
sensitivity to disorder, leading to localization-induced non-Fermi
liquid behavior. Our results demonstrate that a well defined Griffiths
phase can exist, which is not restricted to the vicinity of any
magnetic ordering and yet seems to be consistent with most puzzling
features of disordered heavy fermion systems and Kondo alloys.

We acknowledge useful discussions with M. C. Aronson, D. L. Cox,
G. Kotliar, D. E. MacLaughlin, A. J. Millis, and G. Zarand.  This work
was supported by FAPESP and CNPq (EM), NSF grant DMR-9974311 and the
Alfred P. Sloan Foundation (VD).


\begin{references}
\bibitem{tk} In the Bethe lattice, impurity-related quantities such
as the Kondo temperature $T_K$ actually come from solving an action
like~(\ref{seff}) with $z$ instead of $z-1$ neighbors. This leads to
quantitative but not qualitative changes (see ref.~\cite{sdmft} for a
discussion).
\end{references}
\end{document}